\documentclass[12pt]{article}%
\usepackage{amsmath}
\usepackage{amsfonts}
\usepackage{amssymb}
\usepackage{graphicx}%
\newenvironment{proof}[1][Proof]{\noindent\textbf{#1.} }{\ \rule{0.5em}{0.5em}}
\begin{document}

\author{{\small V. V. Fern\'{a}ndez}$^{{\footnotesize 1}}${\small , A. M.
Moya}$^{{\footnotesize 1}}${\small , E. Notte-Cuello}$^{{\footnotesize 2}}%
${\small \ and W. A. Rodrigues Jr.}$^{{\footnotesize 1}}${\small . }\\$^{{\footnotesize 1}}\hspace{-0.1cm}${\footnotesize Institute of Mathematics,
Statistics and Scientific Computation}\\{\footnotesize IMECC-UNICAMP CP 6065}\\{\footnotesize 13083-859 Campinas, SP, Brazil}\\$^{{\footnotesize 2}}${\small Departamento de} {\small Matem\'{a}ticas,}\\{\small Universidad de La Serena}\\{\small Av. Cisternas 1200, La Serena-Chile}\\{\small e-mail:} {\small walrod@ime.unicamp.br and enotte@userena.cl }}
\title{Parallelism Structure on a Smooth Manifold}
\maketitle

\begin{abstract}
Using the theory of extensors developed in a previous paper \emph{\ }we
present a theory of the parallelism structure on arbitrary smooth manifold.
Two kinds of Cartan connection operators are introduced and both appear in
intrinsic versions (i.e., frame independent) of the first and second Cartan
structure equations. Also, the concept of deformed parallelism structures and
relative parallelism structures which play important role in the understanding
of geometrical theories of the gravitational field are investigated.

\end{abstract}

\newpage

\section{Introduction}

In this article using mainly the algebraic tools developed in
\cite{fmcr1,qr07} we present \ a theory of a general parallelism structure on
an arbitrary real differentiable manifold $M$ of dimension $n$. Section 2
recalls briefly the concepts of covariant derivatives of vector, form and
extensor fields. Section 3 introduce \textit{two} kinds of Cartan connection
operators, namely the \textit{plus} and \textit{minus} Cartan connections.
With these conceptions we show in Section 4 an intrinsic version of Cartan
first structure equation for a biform valued (1,2) extensor field $\Theta$
(torsion) which involves the minus connection. In section 5 we show an
intrinsic Cartan second structure equation for a (1 vector and 1 form) biform
valued curvature extensor $\Omega$ where both the plus and minus connection
naturally appears! In Section 7 we study the concept of a symmetric
parallelism structure and in Section 8 we present the important concept of
deformed parallelism structures. In Section 8 we introduce the concept of
relative parallelism structure. These concepts play an important role in any
deep study of geometric theories of the gravitational field. In Section 10 we
present our conclusions. The present paper and the sequel ones
\cite{fmcr3,fmcr4} constitutes a valuable and simplifying improvement over the
presentation given in \cite{fmr107,fmr207,fmr307,fmr407} which uses only the
geometric algebra of multivector fields and $%
{\displaystyle\bigwedge}
TM$ valued extensor fields.

\section{Parallelism Structure}

Let $U$ be an open set of the smooth manifold $M$ (i.e., $U\subseteq M$). The
set of smooth\footnote{Smooth in this paper means i.e., $\mathcal{C}^{\infty}%
$-differentiable or at least enough differentiable in order for our statements
to hold.} scalar fields on $U,$ as well-known, has a natural structure of
\emph{ring} (\emph{with identity}), and it will be denoted by $\mathcal{S}%
(U).$ The set of smooth vector fields on $U$ and the set of smooth form fields
on $U,$ as well-known, have natural structure of \emph{modules over}
$\mathcal{S}(U).$

A smooth $2$ vector variables vector operator field on $U,$
\[
\Gamma:\mathcal{V}(U)\times\mathcal{V}(U)\longrightarrow\mathcal{V}(U),
\]
such that it satisfies the following axioms:

\textbf{i. }for all $f,g\in\mathcal{S}(U)$ and $a,b,v\in\mathcal{V}(U)$%
\begin{equation}
\Gamma(fa+gb,v)=f\Gamma(a,v)+g\Gamma(b,v), \label{PS1}%
\end{equation}

\textbf{ii.} for all $f,g\in\mathcal{S}(U)$ and $a,v,w\in\mathcal{V}(U)$%
\begin{equation}
\Gamma(a,fv+gw)=(af)v+(ag)w+f\Gamma(a,v)+g\Gamma(a,w), \label{PS2}%
\end{equation}
is called a \emph{connection }on $U.$

The behavior of $\Gamma$ with respect to the first variable will be called
\emph{strong linearity}, and the behavior of $\Gamma$ with respect to the
second variable will be called \emph{quasi linearity}.

The algebraic pair $\left\langle U,\Gamma\right\rangle ,$ where $U$ is an open
set of the smooth manifold $M$ and $\Gamma$ is a connection on $U,$ will be
called a \emph{parallelism structure} on $U.$

\subsection{Covariant Derivative of Vector and Form Fields}

Let $\left\langle U,\Gamma\right\rangle $ be a parallelism structure on $U.$
Let us take $a\in\mathcal{V}(U).$ The $a$\emph{-Directional Covariant
Derivative} ($a$\emph{-DCD})\emph{\ }of a smooth vector field on $U,$
associated with $\left\langle U,\Gamma\right\rangle ,$ is the mapping
\[
\mathcal{V}(U)\ni v\longmapsto\nabla_{a}v\in\mathcal{V}(U)
\]
such that
\begin{equation}
\nabla_{a}v=\Gamma(a,v).\label{CDV1}%
\end{equation}

The $a$\emph{-DCD }of a smooth form field on $U,$ is the mapping
\[
V^{\ast}(U)\ni\omega\longmapsto\nabla_{a}\omega\in V^{\ast}(U)
\]
such that for every $v\in\mathcal{V}(U)$
\begin{equation}
\nabla_{a}\omega(v)=a\omega(v)-\omega(\nabla_{a}v). \label{CDF1}%
\end{equation}

Eq.(\ref{PS1}) and Eq.(\ref{PS2}), according with Eq.(\ref{CDV1}), imply two
basic properties for the covariant derivative of smooth vector fields:

\begin{itemize}
\item For all $f\in S(U),$ and $a,b,v\in\mathcal{V}(U)$
\begin{align}
\nabla_{a+b}v  &  =\nabla_{a}v+\nabla_{b}v,\nonumber\\
\nabla_{fa}v  &  =f\nabla_{a}v. \label{CDV2}%
\end{align}

\item For all $f\in S(U),$ and $a,v,w\in\mathcal{V}(U)$%
\begin{align}
\nabla_{a}(v+w)  &  =\nabla_{a}v+\nabla_{a}w,\nonumber\\
\nabla_{a}(fv)  &  =(af)v+f\nabla_{a}v. \label{CDV3}%
\end{align}

\end{itemize}

From Eq.(\ref{CDF1}), using the strong linearity of the smooth form fields and
recalling the Leibniz property of the $a$-directional derivative of the smooth
scalar fields, it follows two basic properties for the covariant derivative of
the smooth form fields:

\begin{itemize}
\item For all $f\in S(U),$ and $a,b\in\mathcal{V}(U),$ and $\omega
\in\mathcal{V}^{\ast}(U)$
\begin{align}
\nabla_{a+b}\omega &  =\nabla_{a}\omega+\nabla_{b}\omega,\nonumber\\
\nabla_{fa}\omega &  =f\nabla_{a}\omega. \label{CDF2}%
\end{align}

\item For all $f\in S(U),$ and $a\in\mathcal{V}(U)$ and $\omega,\sigma
\in\mathcal{V}^{\ast}(U)$%
\begin{align}
\nabla_{a}(\omega+\sigma)  &  =\nabla_{a}\omega+\nabla_{a}\sigma,\nonumber\\
\nabla_{a}(f\omega)  &  =(af)\omega+f\nabla_{a}\omega. \label{CVF3}%
\end{align}

\end{itemize}

\subsection{Covariant Derivative of Elementary Extensor Fields}

We recall that there exist exactly two types of \emph{elementary }extensors
over a real vector space $V$ of finite dimension\footnote{$V$ could be also a
module over a ring.}, i.e., the linear mappings of the type
\begin{align}
\underset{k\text{-copies}}{\underbrace{V\times\cdots\times V}}\times
\underset{l\text{-copies}}{\underbrace{V^{\ast}\times\cdots\times V^{\ast}}}
&  \ni(v_{1},\ldots,v_{k},\omega^{1},\ldots,\omega^{l})\nonumber\\
&  \longmapsto\tau(v_{1},\ldots,v_{k},\omega^{1},\ldots,\omega^{l})\in V
\label{EEF1}%
\end{align}
called a $k$\emph{-covariant and }$l$\emph{-contravariant vector extensor
over} $V$, and the linear mappings of the type
\begin{align}
\underset{k\text{-copies}}{\underbrace{V\times\cdots\times V}}\times
\underset{l\text{-copies}}{\underbrace{V^{\ast}\times\cdots\times V^{\ast}}}
&  \ni(v_{1},\ldots,v_{k},\omega^{1},\ldots,\omega^{l})\nonumber\\
&  \longmapsto\tau(v_{1},\ldots,v_{k},\omega^{1},\ldots,\omega^{l})\in
V^{\ast} \label{EEF2}%
\end{align}
called a $k$\emph{-covariant and }$l$\emph{-contravariant form extensor over}
$V$.

The set of each of them are denoted by $\left.  \overset{}{ext}\right.
_{k}^{l}(V)$ and $\left.  \overset{\ast}{ext}\right.  _{k}^{l}(V),$\ and have
natural structures of real vector spaces and moreover
\begin{equation}
\dim\left.  \overset{}{ext}\right.  _{k}^{l}(V)=\dim\left.  \overset{\ast
}{ext}\right.  _{k}^{l}(V)=n^{k+l+1}. \label{EEF3}%
\end{equation}

Let $U$ be an open set of the manifold $M$ (i.e., $U\subseteq M$). A mapping
\begin{equation}
U\ni p\longmapsto\tau_{(p)}\in\left.  \overset{}{ext}\right.  _{k}^{l}(T_{p}M)
\label{EEF4}%
\end{equation}
is called a $k$\emph{-covariant and }$l$\emph{-contravariant vector extensor
field }on $U.$

Such a extensor field $\tau$ is said to be \emph{smooth} if and only if for
all $v_{1},\ldots,v_{k}\in\mathcal{V}(U),$ and $\omega^{1},\ldots,\omega
^{l}\in\mathcal{V}^{\ast}(U)$ the mapping
\begin{equation}
U\ni p\longmapsto\tau_{(p)}(v_{1(p)},\ldots,v_{k(p)},\omega_{(p)}^{1}%
,\ldots,\omega_{(p)}^{l})\in T_{p}M \label{EEF5a}%
\end{equation}
is a\ smooth vector field on $U.$

A mapping%
\begin{equation}
U\ni p\longmapsto\tau_{(p)}\in\left.  \overset{\ast}{ext}\right.  _{k}%
^{l}(T_{p}M) \label{EEF6}%
\end{equation}
is called a $k$\emph{-covariant and }$l$\emph{-contravariant form extensor
field }on $U.$

Such a extensor field $\tau$ is said to be \emph{smooth} if and only if for
all $v_{1},\ldots,v_{k}\in\mathcal{V}(U),$ and $\omega^{1},\ldots,\omega
^{l}\in\mathcal{V}^{\ast}(U)$ the mapping
\begin{equation}
U\ni p\longmapsto\tau_{(p)}(v_{1(p)},\ldots,v_{k(p)},\omega_{(p)}^{1}%
,\ldots,\omega_{(p)}^{l})\in T_{p}^{\ast}M \label{EEF7}%
\end{equation}
is a smooth form field on $U.$

The sets of these smooth extensor field on $U$ are denoted by $\left.
\overset{}{ext}\right.  _{k}^{l}\mathcal{V}(U)$ and $\left.  \overset{\ast
}{ext}\right.  _{k}^{l}\mathcal{V}(U),$ since according with the definitions
of smoothness for extensor fields on $U$ as given above, each of smooth
extensor field on $U$ could be properly seen as some type of \emph{extensor
over} $\mathcal{V}(U).$

Let $\left\langle U,\Gamma\right\rangle $ be a parallelism structure on $U$.
Take $a\in\mathcal{V}(U)$.

The\emph{\ }$a$\emph{-DCD} \emph{of a smooth }$k$\emph{-covariant and }%
$l$\emph{-contravariant vector }(\emph{or form})\emph{\ extensor field} on
$U,$ and associated with $\left\langle U,\Gamma\right\rangle ,$ are defined
by
\[
\left.  \overset{}{ext}\right.  _{k}^{l}\mathcal{V}(U)\ni\tau\longmapsto
\nabla_{a}\tau\in\left.  \overset{}{ext}\right.  _{k}^{l}\mathcal{V}(U)\text{
and }\left.  \overset{\ast}{ext}\right.  _{k}^{l}\mathcal{V}(U)\ni
\tau\longmapsto\nabla_{a}\tau\in\left.  \overset{\ast}{ext}\right.  _{k}%
^{l}\mathcal{V}(U),
\]
respectively, such that for every $v_{1},\ldots,v_{k}\in\mathcal{V}(U),$ and
$\omega^{1},\ldots,\omega^{l}\in V^{\ast}(U):$%
\begin{align}
\nabla_{a}\tau(v_{1},\ldots,v_{k},\omega^{1},\ldots,\omega^{l}) &  =\nabla
_{a}(\tau(v_{1},\ldots,v_{k},\omega^{1},\ldots,\omega^{l}))\nonumber\\
&  -\tau(\nabla_{a}v_{1},\ldots,v_{k},\omega^{1},\ldots,\omega^{l}%
)-\cdots\nonumber\\
&  -\tau(v_{1},\ldots,\nabla_{a}v_{k},\omega^{1},\ldots,\omega^{l})\nonumber\\
&  -\tau(v_{1},\ldots,v_{k},\nabla_{a}\omega^{1},\ldots,\omega^{l}%
)-\cdots\nonumber\\
&  -\tau(v_{1},\ldots,v_{k},\omega^{1},\ldots,\nabla_{a}\omega^{l}%
).\label{EEF8}%
\end{align}

We present two basic properties for the covariant derivative of smooth
extensor fields:

\begin{itemize}
\item For $f\in\mathcal{S}(U),$ and $a,b\in\mathcal{V}(U),\mathcal{\ }$and
$\tau\in\left.  \overset{}{ext}\right.  _{k}^{l}\mathcal{V}(U)$ (or $\tau
\in\left.  \overset{\ast}{ext}\right.  _{k}^{l}\mathcal{V}(U)$)
\begin{align}
\nabla_{a+b}\tau &  =\nabla_{a}\tau+\nabla_{b}\tau\nonumber\\
\nabla_{fa}\tau &  =f\nabla_{a}\tau. \label{EEF9}%
\end{align}

\item For $f\in\mathcal{S}(U),$ and $a\in\mathcal{V}(U),\mathcal{\ }$and
$\tau,\sigma\in\left.  \overset{}{ext}\right.  _{k}^{l}\mathcal{V}(U)$ (or
$\tau,\sigma\in\left.  \overset{\ast}{ext}\right.  _{k}^{l}\mathcal{V}(U)$)
\begin{align}
\nabla_{a}(\tau+\sigma)  &  =\nabla_{a}\tau+\nabla_{a}\sigma,\nonumber\\
\nabla_{a}(f\tau)  &  =(af)\tau+f\nabla_{a}\tau. \label{EEF10}%
\end{align}

\end{itemize}

\section{Cartan Connections}

The smooth $1$\emph{\ vector and }$1$\emph{\ form variables form operator
field }on $U,$%
\[
\Gamma^{+}:\mathcal{V}(U)\times\mathcal{V}^{\ast}(U)\longrightarrow V^{\ast
}(U)
\]
such that%
\begin{equation}
\Gamma^{+}(v,\omega)=\left\langle \omega,\nabla_{e_{\sigma}}v\right\rangle
\varepsilon^{\sigma},\label{CC1}%
\end{equation}
where $\left\{  e_{\mu},\varepsilon^{\mu}\right\}  $ is any pair of dual frame
fields on $V\supseteq U,$ will be called \emph{plus Cartan connection} on $U.$
We emphasis that $\Gamma^{+}$, as a mapping associated with $\Gamma$ is
well-defined, since the form field $\Gamma^{+}(v,\omega)$ does not depend on
the pair $\left\{  e_{\mu},\varepsilon^{\mu}\right\}  $ which is chosen for
calculating it.

The plus Cartan connection has the basic properties:

\begin{itemize}
\item For all $f\in\mathcal{S}(U)$ and $v,w\in\mathcal{V}(U)$ and $\omega
\in\mathcal{V}^{\ast}(U)$%
\begin{align}
\Gamma^{+}(v+w,\omega)  &  =\Gamma^{+}(v,\omega)+\Gamma^{+}(w,\omega
),\nonumber\\
\Gamma^{+}(fv,\omega)  &  =\left\langle \omega,v\right\rangle df+f\Gamma
^{+}(v,\omega). \label{CC2}%
\end{align}
We note that $\Gamma^{+}$ has \emph{quasi linearity} with respect to the first variable.

\item For all $f\in S(U)$ and $\omega,\sigma\in\mathcal{V}^{\ast}(U)$ and
$v\in\mathcal{V}(U)$%
\begin{align}
\Gamma^{+}(v,\omega+\sigma)  &  =\Gamma^{+}(v,\omega)+\Gamma^{+}%
(v,\sigma),\nonumber\\
\Gamma^{+}(v,f\omega)  &  =f\Gamma^{+}(v,\omega). \label{CC3}%
\end{align}
We note that $\Gamma^{+}$ has \emph{strong linearity} with respect to the
second variable.

\item $\Gamma^{+}$ has a kind of \emph{inversion property}
\end{itemize}

\begin{equation}
\left\langle \Gamma^{+}(v,\omega),a\right\rangle =\left\langle \omega
,\nabla_{a}v\right\rangle . \label{CC4}%
\end{equation}
Then,%
\begin{equation}
\nabla_{a}v=\left\langle \Gamma^{+}(v,\varepsilon^{\sigma}),a\right\rangle
e_{\sigma}, \label{CC5}%
\end{equation}
which gives $\nabla_{a}v$ (the $a$-\emph{DCD} of a smooth vector field $v$) in
terms of $\Gamma^{+}(v,\omega).$ It should be noted that the strong linearity
of $\Gamma^{+}$ with respect to the second variable is essential in the
realization of the \emph{frame field independent character} on the right side
of Eq.(\ref{CC5}), what is logically necessary for consistency.

The smooth $1$\emph{\ vector and }$1$\emph{\ form variables form operator
field }on $U,$%
\[
\Gamma^{-}:\mathcal{V}(U)\times\mathcal{V}^{\ast}(U)\longrightarrow V^{\ast
}(U)
\]
such that%
\begin{equation}
\Gamma^{-}(v,\omega)=\left\langle \nabla_{e_{\sigma}}\omega,v\right\rangle
\varepsilon^{\sigma},\label{CC6}%
\end{equation}
where as before $\left\{  e_{\mu},\varepsilon^{\mu}\right\}  $ is any pair of
dual frame field on $V\supseteq U$, will be called \emph{minus Cartan
connection} on $U.$ We emphasis that $\Gamma^{-}$, as mapping associated with
$\Gamma,$ is also well-defined.

The minus Cartan connection has the basic properties:

\begin{itemize}
\item For all $f\in\mathcal{S}(U)$ and $v,w\in\mathcal{V}(U)$ and $\omega
\in\mathcal{V}^{\ast}(U)$%
\begin{align}
\Gamma^{-}(v+w,\omega)  &  =\Gamma^{-}(v,\omega)+\Gamma^{-}(w,\omega
),\nonumber\\
\Gamma^{-}(fv,\omega)  &  =f\Gamma^{-}(v,\omega). \label{CC7}%
\end{align}
That is, $\Gamma^{-}$ has \emph{strong linearity} with respect to the first variable.

\item For all $f\in S(U)$ and $\omega,\sigma\in\mathcal{V}^{\ast}(U)$ and
$v\in\mathcal{V}(U)$%
\begin{align}
\Gamma^{-}(v,\omega+\sigma)  &  =\Gamma^{-}(v,\omega)+\Gamma^{-}%
(v,\sigma),\nonumber\\
\Gamma^{-}(v,f\omega)  &  =\left\langle \omega,v\right\rangle df+f\Gamma
^{-}(v,\omega). \label{CC8}%
\end{align}
That is, $\Gamma^{-}$ has \emph{quasi linearity} with respect to the second variable.

\item $\Gamma^{-}$ has also a kind of \emph{inversion property}, i.e.,
\end{itemize}

\begin{equation}
\left\langle \Gamma^{-}(v,\omega),a\right\rangle =\left\langle \nabla
_{a}\omega,v\right\rangle . \label{CC9}%
\end{equation}
Then,%
\begin{equation}
\nabla_{a}\omega=\left\langle \Gamma^{-}(e_{\sigma},\omega),a\right\rangle
\varepsilon^{\sigma}. \label{CC10}%
\end{equation}
which gives $\nabla_{a}\omega$ (the $a$-\emph{DCD} of a smooth form field
$\omega$) in terms of $\Gamma^{-}(v,\omega)$. We emphasize that the strong
linearity of $\Gamma^{-}$ with respect to the first variable is essential for
logical consistency of Eq.(\ref{CC10}).

The relationship between $\Gamma^{+}$ and $\Gamma^{-}$ is given by the
noticeable property

\begin{itemize}
\item
\begin{equation}
\Gamma^{+}(v,\omega)+\Gamma^{-}(v,\omega)=d\left\langle \omega,v\right\rangle
. \label{CC11}%
\end{equation}

\end{itemize}

The well-known \emph{Cartan connections forms} associated with some pair of
dual frame fields $\left\{  e_{\mu},\varepsilon^{\mu}\right\}  $, here denoted
by $\gamma_{\mu}^{\nu},$ are just the values of $\Gamma^{+}$ and $\Gamma^{-}$
evaluated for $(e_{\mu},\varepsilon^{\mu})$, i.e.,%
\begin{equation}
\gamma_{\mu}^{\nu}=\Gamma^{+}(e_{\mu},\varepsilon^{\nu})=-\Gamma^{-}(e_{\mu
},\varepsilon^{\nu}).\label{CC12}%
\end{equation}

\section{Torsion}

Let $\left\langle U,\Gamma\right\rangle $ be a parallelism structure on $U.$
The smooth $2$\emph{-covariant vector extensor field} on $U,$ defined by
\[
\mathcal{V}(U)\times\mathcal{V}(U)\ni(a,b)\longmapsto\tau(a,b)\in
\mathcal{V}(U)
\]
such that
\begin{equation}
\tau(a,b)=\nabla_{a}b-\nabla_{b}a-\left[  a,b\right]  , \label{TF1}%
\end{equation}
will be called the \emph{fundamental torsion extensor field} of $\left\langle
U,\Gamma\right\rangle $.

\begin{itemize}
\item $\tau$ is skew-symmetric, i.e.,
\begin{equation}
\tau(b,a)=-\tau(a,b).\label{TF2}%
\end{equation}
Accordingly, there exists a smooth $(2,1)$\emph{-extensor field }on $U,$
defined by%
\[%
{\displaystyle\bigwedge\nolimits^{2}}
\mathcal{V}(U)\ni X^{2}\longmapsto\mathcal{T}(X^{2})\in\mathcal{V}(U)
\]
such that%
\begin{equation}
\mathcal{T}(X^{2})=\frac{1}{2}\left\langle \varepsilon^{\mu}\wedge
\varepsilon^{\nu},X^{2}\right\rangle \tau(e_{\mu},e_{\nu}),\label{TF3}%
\end{equation}
where $\left\{  e_{\mu},\varepsilon^{\mu}\right\}  $ is any pair of dual frame
fields on $V\supseteq U.$ It should be emphasized that $\mathcal{T}$, as
extensor field associated with $\tau$, is well-defined since the vector field
$\mathcal{T}(X^{2})$ does not depend on the choice of $\left\{  e_{\mu
},\varepsilon^{\mu}\right\}  .$

\item Such a \emph{torsion extensor field} $\mathcal{T}$ $\ $has the basic
property%
\begin{equation}
\mathcal{T}(a\wedge b)=\tau(a,b). \label{TF4}%
\end{equation}

\end{itemize}

\begin{proof}
A straightforward calculation yields%
\begin{align*}
\mathcal{T}(a\wedge b)  &  =\frac{1}{2}\left\langle \varepsilon^{\mu}%
\wedge\varepsilon^{\nu},a\wedge b\right\rangle \tau(e_{\mu},e_{\nu})=\frac
{1}{2}\det\left(
\begin{array}
[c]{ll}%
\varepsilon^{\mu}(a) & \varepsilon^{\mu}(b)\\
\varepsilon^{\nu}(a) & \varepsilon^{\nu}(b)
\end{array}
\right)  \tau(e_{\mu},e_{\nu})\\
&  =\frac{1}{2}\left(  \varepsilon^{\mu}(a)\varepsilon^{\nu}(b)-\varepsilon
^{\mu}(b)\varepsilon^{\nu}(a)\right)  \tau(e_{\mu},e_{\nu})=\frac{1}{2}\left(
\tau(a,b)-\tau(b,a)\right)  ,
\end{align*}
and, by taking into account Eq.(\ref{TF2}), the expected result follows.
\end{proof}

It should be remarked that the well known \emph{torsion tensor field }is just
given by%
\[
\mathcal{V}(U)\times\mathcal{V}(U)\times\mathcal{V}^{\ast}(U)\ni
(a,b,\omega)\longmapsto T(a,b,\omega)\in\mathcal{S}(U)
\]
such that%
\begin{equation}
T(a,b,\omega)=\left\langle \omega,\tau(a,b)\right\rangle . \label{TF5}%
\end{equation}

Note that it is possible to get $\tau$ in terms of $T,$ i.e.,%
\begin{equation}
\tau(a,b)=T(a,b,\varepsilon^{\mu})e_{\mu}, \label{TF6}%
\end{equation}
where $\left\{  e_{\mu},\varepsilon^{\mu}\right\}  $ is any pair of dual frame
fields on $V\supseteq U.$

It is also possible to introduce a \emph{third torsion extensor field} for
$\left\langle U,\Gamma\right\rangle $ by defining the smooth $(1,2)$%
\emph{-extensor field }on $U$,%
\[
V^{\ast}(U)\ni\omega\longmapsto\Theta(\omega)\in%
{\displaystyle\bigwedge\nolimits^{2}}
V^{\ast}(U),
\]
such that
\begin{equation}
\Theta(\omega)=\frac{1}{2}\left\langle \omega,\tau(e_{\mu},e_{\nu
})\right\rangle \varepsilon^{\mu}\wedge\varepsilon^{\nu}, \label{TF7}%
\end{equation}
where $\left\{  e_{\mu},\varepsilon^{\mu}\right\}  $ is any pair of dual frame
fields on $V\supseteq U$. We call $\Theta$ the \emph{Cartan torsion extensor
field} of $\left\langle U,\Gamma\right\rangle .$

The choice of this name is seem to be appropriate once we note that the
so-called \emph{Cartan torsion biforms} associated with some pair of dual
frame fields $\left\{  e_{\mu},\varepsilon^{\mu}\right\}  ,$ usually denoted
by $\Theta^{\nu},$ are just the values of $\Theta$ evaluate  for
$\varepsilon^{\nu}$, i.e.,%

\[
\Theta^{\nu}=\Theta(\varepsilon^{\nu})=\frac{1}{2}\left\langle \varepsilon
^{\nu},\tau(e_{\alpha},e_{\beta})\right\rangle \varepsilon^{\alpha}%
\wedge\varepsilon^{\beta}=\frac{1}{2}T(e_{\alpha},e_{\beta},\varepsilon^{\nu
})\varepsilon^{\alpha}\wedge\varepsilon^{\beta}=\frac{1}{2}T_{\alpha\beta
}^{\nu}\varepsilon^{\alpha}\wedge\varepsilon^{\beta}.
\]

Finally, we present and prove a noticeable property:

\begin{itemize}
\item $\mathcal{T}$ and $\Theta$ are \emph{duality adjoint} to each other,
i.e.,%
\[
\mathcal{T}^{\triangle}=\Theta\text{ and }\Theta^{\triangle}=\mathcal{T}.
\]

\end{itemize}

\begin{proof}
Take $X^{2}\in%
{\displaystyle\bigwedge\nolimits^{2}}
\mathcal{V}(U)$ and $\omega\in\mathcal{V}^{\ast}(U).$ We must prove that%
\[
\left\langle \omega,\mathcal{T}(X^{2})\right\rangle =\left\langle
\Theta(\omega),X^{2}\right\rangle .
\]
Using Eq.(\ref{TF3}) and Eq.(\ref{TF7}), we can write that%
\begin{align*}
\left\langle \omega,\mathcal{T}(X^{2})\right\rangle  &  =\left\langle
\omega,\frac{1}{2}\left\langle \varepsilon^{\mu}\wedge\varepsilon^{\nu}%
,X^{2}\right\rangle \tau(e_{\mu},e_{\nu})\right\rangle =\frac{1}%
{2}\left\langle \varepsilon^{\mu}\wedge\varepsilon^{\nu},X^{2}\right\rangle
\left\langle \omega,\tau(e_{\mu},e_{\nu})\right\rangle \\
&  =\left\langle \frac{1}{2}\left\langle \omega,\tau(e_{\mu},e_{\nu
})\right\rangle \varepsilon^{\mu}\wedge\varepsilon^{\nu},X^{2}\right\rangle
=\left\langle \Theta(\omega),X^{2}\right\rangle ,
\end{align*}
which proves our statement.
\end{proof}

\subsection{Cartan First Equation}

\begin{itemize}
\item Let $\left\langle U,\Gamma\right\rangle $ be a parallelism structure on
$U$. Take any pair of dual frame fields $\left\{  e_{\mu},\varepsilon^{\mu
}\right\}  $ on $V\supseteq U$. We have
\begin{equation}
\Theta(\omega)=d\omega-\Gamma^{-}(e_{\sigma},\omega)\wedge\varepsilon^{\sigma}
\label{TF8}%
\end{equation}

\end{itemize}

\begin{proof}
A straightforward calculation yields%
\begin{align*}
\Theta(\omega)  &  =\frac{1}{2}\left\langle \omega,\nabla_{e_{\mu}}e_{\nu
}-\nabla_{e_{\nu}}e_{\mu}-\left[  e_{\mu},e_{\nu}\right]  \right\rangle
\varepsilon^{\mu}\wedge\varepsilon^{\nu}\\
&  =\left\langle \omega,\nabla_{e_{\mu}}e_{\nu}\right\rangle \varepsilon^{\mu
}\wedge\varepsilon^{\nu}-\frac{1}{2}\left\langle \omega,\left[  e_{\mu}%
,e_{\nu}\right]  \right\rangle \varepsilon^{\mu}\wedge\varepsilon^{\nu},
\end{align*}
but, by recalling the identity%
\[
d\omega=d\omega(e_{\nu})\wedge\varepsilon^{\nu}-\frac{1}{2}\left\langle
\omega,\left[  e_{\mu},e_{\nu}\right]  \right\rangle \varepsilon^{\mu}%
\wedge\varepsilon^{\nu}%
\]
valid for smooth form fields we get%
\[
\Theta(\omega)=\Gamma^{+}(e_{\nu},\omega)\wedge\varepsilon^{\nu}+\left(
d\omega-d\omega(e_{\nu}\right)  \wedge\varepsilon^{\nu}),
\]
from where using Eq.(\ref{CC11}) the expected result follows.
\end{proof}

We emphasis that Eq.(\ref{TF8}) is the \emph{frame field independent version
}of the so-called \emph{Cartan first structure equation}. In fact, if we
choose some pair of dual frame fields $\left\{  e_{\mu},\varepsilon^{\mu
}\right\}  ,$ we can write
\[
\Theta(\varepsilon^{\nu})=d\varepsilon^{\nu}-\Gamma^{-}(e_{\sigma}%
,\varepsilon^{\nu})\wedge\varepsilon^{\sigma},
\]
i.e.,
\begin{equation}
\Theta^{\nu}=d\varepsilon^{\nu}+\gamma_{\sigma}^{\nu}\wedge\varepsilon
^{\sigma}. \label{TF9}%
\end{equation}

What the meaning of the second term in Eq.(\ref{TF8})?\footnote{Note that the
strong linearity of $\Gamma^{-}$ is essential in warranting that $\Gamma
^{-}(e_{\sigma},\omega)\wedge\varepsilon^{\sigma}$ has frame field independent
character, i.e., that it does not depend on the choice of $\left\{  e_{\mu
},\varepsilon^{\mu}\right\}  $.}

A straightforward calculation gives
\begin{align}
\Gamma^{-}(e_{\sigma},\omega)\wedge\varepsilon^{\sigma}  &  =\left\langle
\nabla_{e_{\mu}}\omega,e_{\sigma}\right\rangle \wedge\varepsilon^{\sigma
}=\varepsilon^{\mu}\wedge\nabla_{e_{\mu}}\omega,\nonumber\\
\Gamma^{-}(e_{\sigma},\omega)\wedge\varepsilon^{\sigma}  &  =\nabla
\wedge\omega. \label{TF10}%
\end{align}
Thus, the second term in Eq.(\ref{TF8}) is just the\cite{rodoliv2006}
\emph{covariant curl} ($\nabla\wedge\omega$) of the smooth form field $\omega
$. When $\Theta=0$ we have the identity%
\begin{equation}
d\omega=\nabla\wedge\omega.
\end{equation}

\section{Curvature}

Let $\left\langle U,\Gamma\right\rangle $\ be a parallelism structure on $U$.
The smooth $3$\emph{-covariant extensor vector field} on $U$, defined by
\[
\mathcal{V}(U)\times\mathcal{V}(U)\times\mathcal{V}(U)\ni(a,b,c)\longmapsto
\rho(a,b,c)\in\mathcal{V}(U),
\]
such that
\begin{equation}
\rho(a,b,c)=\left[  \nabla_{a},\nabla_{b}\right]  c-\nabla_{\left[
a,b\right]  }c, \label{CF1}%
\end{equation}
will be called the \emph{fundamental curvature extensor field} of
$\left\langle U,\Gamma\right\rangle $.

\begin{itemize}
\item As can be easily verified, $\rho$ is skew-symmetric with respect to the
first and the second variables, i.e.,%
\begin{equation}
\rho(b,a,c)=-\rho(a,b,c). \label{CF2}%
\end{equation}

\end{itemize}

Thus, there exists a smooth $1$\emph{\ bivector and }$1$\emph{\ vector
variables vector extensor field }on $U,$ defined by%
\[%
{\displaystyle\bigwedge\nolimits^{2}}
\mathcal{V}(U)\times\mathcal{V}(U)\ni(X^{2},c)\longmapsto\mathcal{R}%
(X^{2},c)\in\mathcal{V}(U)
\]
such that%
\begin{equation}
\mathcal{R}(X^{2},c)=\frac{1}{2}\left\langle \varepsilon^{\mu}\wedge
\varepsilon^{\nu},X^{2}\right\rangle \rho(e_{\mu},e_{\nu},c),\label{CF3}%
\end{equation}
where $\left\{  e_{\mu},\varepsilon^{\mu}\right\}  $ is any pair of dual frame
fields on $V\supseteq U.$ We note that $\mathcal{R},$ as extensor field
associated with $\rho,$ is well-defined since the vector field $\mathcal{R}%
(X^{2},c)$ does not depend on the choice of $\left\{  e_{\mu},\varepsilon
^{\mu}\right\}  .$

\begin{itemize}
\item The \emph{curvature extensor field} $\mathcal{R}$ has the basic property%
\begin{equation}
\mathcal{R}(a\wedge b,c)=\rho(a,b,c). \label{CF4}%
\end{equation}

\end{itemize}

\begin{proof}
A straightforward calculation gives%
\begin{align*}
\mathcal{R}(a\wedge b,c)  &  =\frac{1}{2}\left\langle \varepsilon^{\mu}%
\wedge\varepsilon^{\nu},a\wedge b\right\rangle \rho(e_{\mu},e_{\nu},c)\\
&  =\frac{1}{2}\det\left(
\begin{array}
[c]{ll}%
\varepsilon^{\mu}(a) & \varepsilon^{\mu}(b)\\
\varepsilon^{\nu}(a) & \varepsilon^{\nu}(b)
\end{array}
\right)  \rho(e_{\mu},e_{\nu},c)\\
&  =\frac{1}{2}\left(  \varepsilon^{\mu}(a)\varepsilon^{\nu}(b)-\varepsilon
^{\mu}(b)\varepsilon^{\nu}(a)\right)  \rho(e_{\mu},e_{\nu},c)\\
&  =\frac{1}{2}\left(  \rho(a,b,c)-\rho(b,a,c)\right)  ,
\end{align*}
and, by using Eq.(\ref{CF2}), we get the expected result.
\end{proof}

We remark that the so-called \emph{curvature tensor field} is just given by%
\[
\mathcal{V}(U)\times\mathcal{V}(U)\times\mathcal{V}(U)\times V^{\ast}%
(U)\ni\left(  a,b,c,\omega\right)  \longmapsto R(a,b,c,\omega)\in
\mathcal{S}(U)
\]
such that
\begin{equation}
R(a,b,c,\omega)=\left\langle \omega,\rho(b,c,a)\right\rangle . \label{CF5}%
\end{equation}

It is possible to get $\rho$ in terms of $R,$ i.e.,%
\begin{equation}
\rho(a,b,c)=R(c,a,b,\varepsilon^{\mu})e_{\mu}, \label{CF6}%
\end{equation}
where $\left\{  e_{\mu},\varepsilon^{\mu}\right\}  $ is any frame field on
$V\supseteq U.$

We note that it is also possible to introduce a \emph{third curvature extensor
field} for $\langle U,\Gamma\rangle$ defining the smooth $1$\emph{\ vector and
}$1$\emph{\ form variables form extensor field} on $U$%
\[
\mathcal{V}(U)\times\mathcal{V}^{\ast}(U)\ni\left(  c,\omega\right)
\longmapsto\Omega\left(  c,\omega\right)  \in%
{\displaystyle\bigwedge\nolimits^{2}}
V^{\ast}(U),
\]
such that%
\begin{equation}
\Omega\left(  c,\omega\right)  =\frac{1}{2}\left\langle \omega,\rho(e_{\mu
},e_{\nu},c)\right\rangle \varepsilon^{\mu}\wedge\varepsilon^{\nu},\label{CF7}%
\end{equation}
where $\left\{  e_{\mu},\varepsilon^{\mu}\right\}  $ is any pair of dual frame
fields on $V\supseteq U$. We call $\Omega$ the \emph{Cartan curvature extensor
field} of $\left\langle U,\Gamma\right\rangle $ since the so-called
\emph{Cartan curvature biforms} associated with some $\left\{  e_{\mu
},\varepsilon^{\mu}\right\}  $, usually denoted by $\Omega_{\mu}^{\nu},$ are
exactly the values of $\Omega$ evaluated for $\left(  e_{\mu},\varepsilon
^{\nu}\right)  $, i.e.%

\[
\Omega_{\mu}^{\nu}:=\Omega\left(  e_{\mu},\varepsilon^{\nu}\right)  =\frac
{1}{2}\left\langle \varepsilon^{\nu},\rho(e_{\alpha},e_{\beta},e_{\mu
})\right\rangle \varepsilon^{\alpha}\wedge\varepsilon^{\beta}=\frac{1}%
{2}R(e_{\mu},e_{\alpha},e_{\beta},\varepsilon^{\nu})\varepsilon^{\alpha}%
\wedge\varepsilon^{\beta}.
\]

Finally, we present and prove a remarkable property:

\begin{itemize}
\item $\mathcal{R}_{c}$ and $\Omega_{c}$ are \emph{duality adjoint }of each
other, i.e.,%
\begin{equation}
\mathcal{R}_{c}^{\triangle}=\Omega_{c}\text{ and }\Omega_{c}^{\triangle
}=\mathcal{R}_{c}. \label{CF8}%
\end{equation}

\end{itemize}

\begin{proof}
Let us take $X^{2}\in%
{\displaystyle\bigwedge\nolimits^{2}}
\mathcal{V}(U)$ and $\omega\in V^{\ast}(U).$ we have to prove that%
\[
\left\langle \omega,\mathcal{R}_{c}(X^{2})\right\rangle =\left\langle
\Omega_{c}(\omega),X^{2}\right\rangle .
\]
By using Eq.(\ref{CF3}) and Eq.(\ref{CF7}), we in fact can write that%
\begin{align*}
\left\langle \omega,\mathcal{R}_{c}(X^{2})\right\rangle  &  =\left\langle
\omega,\frac{1}{2}\left\langle \varepsilon^{\mu}\wedge\varepsilon^{\nu}%
,X^{2}\right\rangle \rho(e_{\mu},e_{\nu},c)\right\rangle \\
&  =\frac{1}{2}\left\langle \varepsilon^{\mu}\wedge\varepsilon^{\nu}%
,X^{2}\right\rangle \left\langle \omega,\rho(e_{\mu},e_{\nu},c)\right\rangle
\\
&  =\left\langle \frac{1}{2}\left\langle \omega,\rho(e_{\mu},e_{\nu
},c)\right\rangle \varepsilon^{\mu}\wedge\varepsilon^{\nu},X^{2}\right\rangle
=\left\langle \Omega_{c}(\omega),X^{2}\right\rangle .
\end{align*}

\end{proof}

\subsection{Cartan Second Equation}

\begin{itemize}
\item Let $\left\langle U,\Gamma\right\rangle $ be a parallelism structure on
$U.$ Let us take any pair of dual frame fields $\left\{  e_{\mu}%
,\varepsilon^{\mu}\right\}  $ on $V\supseteq U.$ We have
\begin{equation}
\Omega(c,\omega)=d\Gamma^{+}(c,\omega)+\Gamma^{+}(c,\varepsilon^{\sigma
})\wedge\Gamma^{-}(e_{\sigma},\omega). \label{CF9}%
\end{equation}

\end{itemize}

\begin{proof}
A straightforward calculation yields%
\begin{align}
\Omega(c,\omega)  &  =\frac{1}{2}\left\langle \omega,\left[  \nabla_{e_{\mu}%
},\nabla_{e_{\nu}}\right]  c-\nabla_{\left[  e_{\mu},e_{\nu}\right]
}c\right\rangle \varepsilon^{\mu}\wedge\varepsilon^{\nu}\nonumber\\
&  =\left\langle \omega,\nabla_{e_{\mu}}\nabla_{e_{\nu}}c\right\rangle
\varepsilon^{\mu}\wedge\varepsilon^{\nu}-\frac{1}{2}\left\langle \omega
,\nabla_{\left[  e_{\mu},e_{\nu}\right]  }c\right\rangle \varepsilon^{\mu
}\wedge\varepsilon^{\nu}, \tag{a}%
\end{align}

By using Eq.(\ref{CC5}), the first term in (a) gives%
\begin{align}
\left\langle \omega,\nabla_{e_{\mu}}\nabla_{e_{\nu}}c\right\rangle
\varepsilon^{\mu}\wedge\varepsilon^{\nu}  &  =\left\langle \omega
,\nabla_{e_{\mu}}\left\langle \Gamma^{+}(c,\varepsilon^{\sigma}),e_{\nu
}\right\rangle e_{\sigma}\right\rangle \varepsilon^{\mu}\wedge\varepsilon
^{\nu}\nonumber\\
&  =\left\langle \omega,e_{\mu}\left\langle \Gamma^{+}(c,\varepsilon^{\sigma
}),e_{\nu}\right\rangle e_{\sigma}\right\rangle \varepsilon^{\mu}%
\wedge\varepsilon^{\nu}\nonumber\\
&  +\left\langle \omega,\left\langle \Gamma^{+}(c,\varepsilon^{\sigma}%
),e_{\nu}\right\rangle \nabla_{e_{\mu}}e_{\sigma}\right\rangle \varepsilon
^{\mu}\wedge\varepsilon^{\nu}\nonumber\\
&  =\left(  e_{\mu}\left\langle \Gamma^{+}(c,\varepsilon^{\sigma}),e_{\nu
}\right\rangle \right)  \left\langle \omega,e_{\sigma}\right\rangle
\varepsilon^{\mu}\wedge\varepsilon^{\nu}\nonumber\\
&  +\left\langle \Gamma^{+}(c,\varepsilon^{\sigma}),e_{\nu}\right\rangle
\left\langle \omega,\nabla_{e_{\mu}}e_{\sigma}\right\rangle \varepsilon^{\mu
}\wedge\varepsilon^{\nu}, \tag{b}%
\end{align}
but, the first term in (b) and the second term in (b) can be written%
\begin{align}
&  \left(  e_{\mu}\left\langle \Gamma^{+}(c,\varepsilon^{\sigma}),e_{\nu
}\right\rangle \right)  \left\langle \omega,e_{\sigma}\right\rangle
\varepsilon^{\mu}\wedge\varepsilon^{\nu}\nonumber\\
&  =d\left\langle \Gamma^{+}(c,\omega),e_{\nu}\right\rangle \wedge
\varepsilon^{\nu}-d\left\langle \omega,e_{\sigma}\right\rangle \wedge
\Gamma^{+}(c,\varepsilon^{\sigma}) \tag{c}%
\end{align}
and%
\begin{equation}
\left\langle \Gamma^{+}(c,\varepsilon^{\sigma}),e_{\nu}\right\rangle
\left\langle \omega,\nabla_{e_{\mu}}e_{\sigma}\right\rangle \varepsilon^{\mu
}\wedge\varepsilon^{\nu}=\Gamma^{+}(e_{\sigma},\omega)\wedge\Gamma
^{+}(c,\varepsilon^{\sigma}). \tag{d}%
\end{equation}

Thus, by putting (c) and (d) into (b), and by taking into account
Eq.(\ref{CC11}), we get%
\begin{equation}
\left\langle \omega,\nabla_{e_{\mu}}\nabla_{e_{\nu}}c\right\rangle
\varepsilon^{\mu}\wedge\varepsilon^{\nu}=d\left\langle \Gamma^{+}%
(c,\omega),e_{\nu}\right\rangle \wedge\varepsilon^{\nu}-\Gamma^{-}(e_{\sigma
},\omega)\wedge\Gamma^{+}(c,\varepsilon^{\sigma}). \tag{e}%
\end{equation}

By using once again Eq.(\ref{CC5}), the second term in (a) gives%
\begin{align}
\frac{1}{2}\left\langle \omega,\nabla_{\left[  e_{\mu},e_{\nu}\right]
}c\right\rangle \varepsilon^{\mu}\wedge\varepsilon^{\nu}  &  =\frac{1}%
{2}\left\langle \omega,\left\langle \Gamma^{+}(c,\varepsilon^{\sigma}),\left[
e_{\mu},e_{\nu}\right]  \right\rangle e_{\sigma}\right\rangle \varepsilon
^{\mu}\wedge\varepsilon^{\nu}\nonumber\\
&  =\frac{1}{2}\left\langle \Gamma^{+}(c,\varepsilon^{\sigma}),\left[  e_{\mu
},e_{\nu}\right]  \right\rangle \left\langle \omega,e_{\sigma}\right\rangle
\varepsilon^{\mu}\wedge\varepsilon^{\nu}\nonumber\\
&  =\frac{1}{2}\left\langle \Gamma^{+}(c,\omega),\left[  e_{\mu},e_{\nu
}\right]  \right\rangle \varepsilon^{\mu}\wedge\varepsilon^{\nu}. \tag{f}%
\end{align}

By putting (e) and (f) into (a), and recalling the identity for smooth form
fields $d\sigma=d\sigma(e_{\nu})\wedge\varepsilon^{\nu}-\frac{1}%
{2}\left\langle \sigma,\left[  e_{\mu},e_{\nu}\right]  \right\rangle
\varepsilon^{\mu}\wedge\varepsilon^{\nu},$ we get the expected result.
\end{proof}

We emphasis that Eq.(\ref{CF9}) is the frame field independent version of the
so-called Cartan second structure equation. Indeed, if we choose some pair of
dual frame fields $\left\{  e_{\mu},\varepsilon^{\mu}\right\}  ,$ we can
write
\[
\Omega(e_{\mu},\varepsilon^{\nu})=d\Gamma^{+}(e_{\mu},\varepsilon^{\nu
})+\Gamma^{+}(e_{\mu},\varepsilon^{\sigma})\wedge\Gamma^{-}(e_{\sigma
},\varepsilon^{\nu}),
\]
i.e.,
\begin{equation}
\Omega_{\mu}^{\nu}=d\gamma_{\mu}^{\nu}+\gamma_{\sigma}^{\nu}\wedge\omega_{\mu
}^{\sigma}. \label{CF10}%
\end{equation}

What is the meaning of the second term in Eq(\ref{CF9})?\footnote{Observe that
the strong linearities of $\Gamma^{+}$ and $\Gamma^{-}$ are essential for
making $\Gamma^{+}(c,\varepsilon^{\sigma})\wedge\Gamma^{-}(e_{\sigma},\omega)$
frame field independent.}

The answer is given by
\begin{align}
\Gamma^{+}(c,\varepsilon^{\sigma})\wedge\Gamma^{-}(e_{\sigma},\omega)  &
=\left\langle \varepsilon^{\sigma},\nabla_{e_{\mu}}c\right\rangle \left\langle
\nabla_{e_{\nu}}\omega,e_{\sigma}\right\rangle \varepsilon^{\mu}%
\wedge\varepsilon^{\nu},\nonumber\\
\Gamma^{+}(c,\varepsilon^{\sigma})\wedge\Gamma^{-}(e_{\sigma},\omega)  &
=\left\langle \nabla_{e_{\nu}}\omega,\nabla_{e_{\mu}}c\right\rangle
\varepsilon^{\mu}\wedge\varepsilon^{\nu}. \label{CF11}%
\end{align}

\section{Symmetric Parallelism Structure}

A parallelism structure $\left\langle U,\Gamma\right\rangle $ is said to be
\emph{symmetric} if and only if for all smooth vector fields $a$ and $b$ it
holds
\begin{equation}
\Gamma(a,b)-\Gamma(b,a)=\left[  a,b\right]  , \label{SPS0}%
\end{equation}
i.e.,
\begin{equation}
\nabla_{a}b-\nabla_{b}a=\left[  a,b\right]  . \label{SPS1}%
\end{equation}

Now, according with Eq.(\ref{TF1}), \ we see that the \emph{condition of
symmetry} is completely equivalent to the \emph{condition of \ null torsion},
i.e.,
\begin{equation}
\tau(a,b)=0. \label{SPS2}%
\end{equation}
So, taking into account Eq.(\ref{TF3}) and Eq.(\ref{TF7}), we also have that
\begin{equation}
\mathcal{T}(X^{2})=0\text{ and }\Theta(\omega)=0. \label{SPS3}%
\end{equation}

We present and prove two noticeable properties for a symmetric parallelism structure.

\begin{itemize}
\item The fundamental curvature field $\rho$ satisfies a \emph{cyclic
property}, i.e.,
\[
\rho(a,b,c)+\rho(b,c,a)+\rho(c,a,b)=0.
\]

\end{itemize}

\begin{proof}
Let us take $a,b,c\in\mathcal{V}(U).$ By using Eq.(\ref{CF1}), we can write
\begin{align}
\rho(a,b,c)  &  =\nabla_{a}\nabla_{b}c-\nabla_{b}\nabla_{a}c-\nabla_{ \left[
a,b\right]  }c,\tag{a}\\
\rho(b,c,a)  &  =\nabla_{b}\nabla_{c}a-\nabla_{c}\nabla_{b}a-\nabla_{ \left[
b,c\right]  }a,\tag{b}\\
\rho(c,a,b)  &  =\nabla_{c}\nabla_{a}b-\nabla_{a}\nabla_{c}b-\nabla_{ \left[
c,a\right]  }b. \tag{c}%
\end{align}

By adding (a), (b) and (c), we\ have
\begin{align}
&  \rho(a,b,c)+\rho(b,c,a)+\rho(c,a,b)\nonumber\\
&  =\nabla_{a}(\nabla_{b}c-\nabla_{c}b)+\nabla_{b}(\nabla_{c}a-\nabla
_{a}c)+\nabla_{c}(\nabla_{a}b-\nabla_{b}a)\nonumber\\
&  -\nabla_{\left[  a,b\right]  }c-\nabla_{\left[  b,c\right]  }%
a-\nabla_{\left[  c,a\right]  }b, \tag{d}%
\end{align}
but, by taking into account Eq.(\ref{SPS1}), we get
\begin{equation}
\rho(a,b,c)+\rho(b,c,a)+\rho(c,a,b)=\left[  a,\left[  b,c\right]  \right]
+\left[  b,\left[  c,a\right]  \right]  +\left[  c,\left[  a,b\right]
\right]  , \tag{e}%
\end{equation}
whence, by recalling the Jacobi identities for the Lie product of smooth
vector fields, the expected result immediately follows.
\end{proof}

\begin{itemize}
\item The fundamental curvature field $\rho$ satisfies the so-called
Bianchi\emph{\ identity}, i.e.,%
\[
\nabla_{w}\rho(a,b,c)+\nabla_{a}\rho(b,w,c)+\nabla_{b}\rho(w,a,c)=0.
\]
Note the cycling of letters: $a,b,w\rightarrow b,w,a\rightarrow w,a,b.$
\end{itemize}

\begin{proof}
Let us take $a,b,c,w\in\mathcal{V}(U).$ By using Eq.(\ref{EEF8}), we have
\begin{align}
&  \nabla_{w}\rho(a,b,c)\nonumber\\
&  =\nabla_{w}(\rho(a,b,c))-\rho(\nabla_{w}a,b,c)-\rho(a,\nabla_{w}%
b,c)-\rho(a,b,\nabla_{w}c),\tag{a}\\
&  \nabla_{a}\rho(b,w,c)\nonumber\\
&  =\nabla_{a}(\rho(b,w,c))-\rho(\nabla_{a}b,w,c)-\rho(b,\nabla_{a}%
w,c)-\rho(b,w,\nabla_{a}c),\tag{b}\\
&  \nabla_{b}\rho(w,a,c)\nonumber\\
&  =\nabla_{b}(\rho(w,a,c))-\rho(\nabla_{b}w,a,c)-\rho(w,\nabla_{b}%
a,c)-\rho(w,a,\nabla_{b}c). \tag{c}%
\end{align}

By adding (a), (b) and (c), and by taking into account Eq.(\ref{CF2}) and
Eq.(\ref{SPS1}), we get
\begin{align}
&  \nabla_{w}\rho(a,b,c)+\nabla_{a}\rho(b,w,c)+\nabla_{b}\rho
(w,a,c)\nonumber\\
&  =\nabla_{w}(\rho(a,b,c))+\nabla_{a}(\rho(b,w,c))+\nabla_{b}(\rho
(w,a,c))\nonumber\\
&  -\rho(\left[  w,a\right]  ,b,c)-\rho(\left[  a,b\right]  ,w,c)-\rho(\left[
b,w\right]  ,a,c)\nonumber\\
&  -\rho(a,b,\nabla_{w}c)-\rho(b,w,\nabla_{a}c)-\rho(w,a,\nabla_{b}c).
\label{d}%
\end{align}

By using Eq.(\ref{CF1}), the first term of (d) can be written
\begin{align}
&  \nabla_{w}(\rho(a,b,c))+\nabla_{a}(\rho(b,w,c))+\nabla_{b}(\rho
(w,a,c))\nonumber\\
&  =\nabla_{w}\left(  \left[  \nabla_{a},\nabla_{b}\right]  c\right)
+\nabla_{a}\left(  \left[  \nabla_{b},\nabla_{w}\right]  c\right)  +\nabla
_{b}\left(  \left[  \nabla_{w},\nabla_{a}\right]  c\right) \nonumber\\
&  -\nabla_{w}\nabla_{\left[  a,b\right]  }c-\nabla_{a}\nabla_{\left[
b,w\right]  }c-\nabla_{b}\nabla_{\left[  w,a\right]  }c. \label{e}%
\end{align}

By using Eq.(\ref{CF1}) and by recalling the so-called Jacobi identity for the
Lie product of smooth vector fields, the second term of (d) can be written
\begin{align}
&  -\rho(\left[  w,a\right]  ,b,c)-\rho(\left[  a,b\right]  ,w,c)-\rho(\left[
b,w\right]  ,a,c)\nonumber\\
&  =-\nabla_{\left[  w,a\right]  }\nabla_{b}c-\nabla_{\left[  a,b\right]
}\nabla_{w}c-\nabla_{\left[  b,w\right]  }\nabla_{a}c\nonumber\\
&  +\nabla_{b}\nabla_{\left[  w,a\right]  }c+\nabla_{w}\nabla_{\left[
a,b\right]  }c+\nabla_{a}\nabla_{\left[  b,w\right]  }c. \tag{f}%
\end{align}
\ By adding (e) and (f), we get
\begin{align}
&  \nabla_{w}(\rho(a,b,c))+\nabla_{a}(\rho(b,w,c))+\nabla_{b}(\rho
(w,a,c))\nonumber\\
&  -\rho(\left[  w,a\right]  ,b,c)-\rho(\left[  a,b\right]  ,w,c)-\rho(\left[
b,w\right]  ,a,c)\nonumber\\
&  =\nabla_{w}\left(  \left[  \nabla_{a},\nabla_{b}\right]  c\right)
+\nabla_{a}\left(  \left[  \nabla_{b},\nabla_{w}\right]  c\right)  +\nabla
_{b}\left(  \left[  \nabla_{w},\nabla_{a}\right]  c\right) \nonumber\\
&  -\nabla_{\left[  w,a\right]  }\nabla_{b}c-\nabla_{\left[  a,b\right]
}\nabla_{w}c-\nabla_{\left[  b,w\right]  }\nabla_{a}c, \label{g}%
\end{align}
now, by sustaining the Eq. (\ref{g}) in the Eq. (\ref{d}), and by using once
again Eq.(\ref{CF1}), we get%
\begin{equation}%
\begin{array}
[c]{l}%
\nabla_{w}(\rho(a,b,c))+\nabla_{a}(\rho(b,w,c))+\nabla_{b}(\rho(w,a,c))\\
=\nabla_{w}\left(  \left[  \nabla_{a},\nabla_{b}\right]  c\right)  +\nabla
_{a}\left(  \left[  \nabla_{b},\nabla_{w}\right]  c\right)  +\nabla_{b}\left(
\left[  \nabla_{w},\nabla_{a}\right]  c\right) \\
-\nabla_{\left[  w,a\right]  }\nabla_{b}c-\nabla_{\left[  a,b\right]  }%
\nabla_{w}c-\nabla_{\left[  b,w\right]  }\nabla_{a}c-\left[  \nabla_{a}%
,\nabla_{b}\right]  \nabla_{w}c+\nabla_{\left[  a,b\right]  }\nabla_{w}c\\
-\left[  \nabla_{b},\nabla_{w}\right]  \nabla_{a}c+\nabla_{\left[  b,w\right]
}\nabla_{a}c-\left[  \nabla_{w},\nabla_{a}\right]  \nabla_{b}c+\nabla_{\left[
w,a\right]  }\nabla_{b}c\\
=\left\{  \left[  \nabla_{w},\left[  \nabla_{a},\nabla_{b}\right]  \right]
+\left[  \nabla_{a},\left[  \nabla_{b},\nabla_{w}\right]  \right]  +\left[
\nabla_{b},\left[  \nabla_{w},\nabla_{a}\right]  \right]  \right\}  c
\end{array}
\label{h}%
\end{equation}
by recalling the so-called Jacobi identities for the Lie product of smooth
vector fields, the expected result immediately follows.
\end{proof}

\section{Deformed Parallelism Structure}

Let $\left\langle U,\Gamma\right\rangle $ be a parallelism structure on $U.$
Let us take an invertible smooth extensor operator field $\lambda$ on
$V\supseteq U,$ i.e., $\lambda:\mathcal{V}(U)\rightarrow\mathcal{V}(U).$ We
can construct another well-defined connection on $U,$ namely $\overset
{\lambda}{\Gamma}$, given by
\[
\mathcal{V}(U)\times\mathcal{V}(U)\ni(a,v)\longmapsto\overset{\lambda}{\Gamma
}(a,v)\in\mathcal{V}(U)
\]
such that
\begin{equation}
\overset{\lambda}{\Gamma}(a,v)=\lambda(\Gamma(a,\lambda^{-1}(v))).
\label{DPS1}%
\end{equation}
$\overset{\lambda}{\Gamma}$ is indeed a connection on $U$ since it satisfies
Eq.(\ref{PS1}) and Eq.(\ref{PS2}). It will be called the $\lambda
$-\emph{deformation} of $\Gamma.$

The parallelism structure $\left\langle U,\overset{\lambda}{\Gamma
}\right\rangle $ is said to be the $\lambda$\emph{-deformation} of
$\left\langle U,\Gamma\right\rangle $.

Let us take $a\in\mathcal{V}(U).$ The $a$\emph{-Directional Covariant
Derivative Operator }($a$-\emph{DCDO}) associated with $\left\langle
U,\overset{\lambda}{\Gamma}\right\rangle ,$ namely $\overset{\lambda}{\nabla
}_{a}$, has the basic properties:

\begin{itemize}
\item For all $v\in\mathcal{V}(U)$%
\begin{equation}
\overset{\lambda}{\nabla}_{a}v=\lambda(\nabla_{a}\lambda^{-1}(v)).
\label{DPS2}%
\end{equation}
It follows from Eq.(\ref{CDV1}) and Eq.(\ref{DPS1}).

\item For all $\omega\in\mathcal{V}^{\ast}(U)$%
\begin{equation}
\overset{\lambda}{\nabla}_{a}\omega=\lambda^{-\bigtriangleup}(\nabla
_{a}\lambda^{\bigtriangleup}(\omega)). \label{DPS4}%
\end{equation}

\end{itemize}

\begin{proof}
Let us take $v\in\mathcal{V}(U).$ By using Eq.(\ref{CDF1}) and Eq.(\ref{DPS2}%
), we have%
\begin{align}
\overset{\lambda}{\nabla}_{a}\omega(v)  &  =a\omega(v)-\omega(\overset
{\lambda}{\nabla}_{a}v)\nonumber\\
&  =a\omega(v)-\omega(\lambda(\nabla_{a}\lambda^{-1}(v)))\nonumber\\
&  =a\left\langle \omega,v\right\rangle -\left\langle \omega,\lambda
(\nabla_{a}\lambda^{-1}(v))\right\rangle , \tag{a}%
\end{align}
but, by recalling the fundamental property of the \emph{duality adjoint} and
by using once again Eq.(\ref{CDF1}), the second term in (a) can be written
\begin{align}
\left\langle \omega,\lambda(\nabla_{a}\lambda^{-1}(v))\right\rangle  &
=\left\langle \lambda^{\bigtriangleup}(\omega),\nabla_{a}\lambda
^{-1}(v)\right\rangle \nonumber\\
&  =a\left\langle \lambda^{\bigtriangleup}(\omega),\lambda^{-1}%
(v)\right\rangle -\left\langle \nabla_{a}\lambda^{\bigtriangleup}%
(\omega),\lambda^{-1}(v)\right\rangle \nonumber\\
&  =a\left\langle \omega,v\right\rangle -\left\langle \lambda^{-\bigtriangleup
}\nabla_{a}\lambda^{\bigtriangleup}(\omega),v\right\rangle , \tag{b}%
\end{align}

Finally, putting (b) into (a), the expected result follows.
\end{proof}

\section{Relative Parallelism Structure}

Let again $\left\{  b_{\mu},\beta^{\mu}\right\}  $ be a pair of dual frame
fields for $U\subseteq M$. Associated with $\left\{  b_{\mu},\beta^{\mu
}\right\}  $ we can construct a well-defined connection on\emph{\ }$U$ given
by the mapping
\[
B:\mathcal{V}(U)\times\mathcal{V}(U)\longrightarrow\mathcal{V}(U),
\]
such that
\begin{equation}
B(a,v)=\left[  a\beta^{\sigma}(v)\right]  b_{\sigma}. \label{RPS1}%
\end{equation}
$B$ \ is called \emph{relative connection }on $U$ with respect to\emph{\ }%
$\left\{  b_{\mu},\beta^{\mu}\right\}  $ (or simply relative connection for short).

The parallelism structure $\left\langle U,B\right\rangle $ will be called
\emph{relative parallelism structure} with respect to $\left\{  b_{\mu}%
,\beta^{\mu}\right\}  .$

The \emph{\ }$a$\emph{-DCDO }induced by the relative connection will be
denoted by $\partial_{a}$. According with Eq.(\ref{CDV1}), the $a$\emph{-DCD}
of a smooth vector field is given by
\begin{equation}
\partial_{a}v=\left[  a\beta^{\sigma}(v)\right]  b_{\sigma}. \label{RPS2}%
\end{equation}

\begin{itemize}
\item Note that $\partial_{a}$ is the unique $a$\emph{-DCDO }which satisfies
the condition
\begin{equation}
\partial_{a}b_{\mu}=0. \label{RPS3}%
\end{equation}

\end{itemize}

From Eq.(\ref{CDF1}) and Eq.(\ref{RPS3}), the $a$\emph{-DCD }of a smooth form
field is given by
\begin{equation}
\partial_{a}\omega=\left[  a\omega(b_{\sigma})\right]  \beta^{\sigma}.
\label{RPS4}%
\end{equation}

Then, it holds also

\begin{itemize}
\item
\begin{equation}
\partial_{a}\beta^{\nu}=0. \label{RPS5}%
\end{equation}

\end{itemize}

The relative parallelism structure has the basic properties:

\begin{itemize}
\item The fundamental torsion extensor field and, the Cartan torsion extensor
field are given by
\begin{align}
\tau(a,b)  &  =\left[  d\beta^{\sigma}(a,b)\right]  b_{\sigma},\label{RPS6}\\
\Theta(\omega)  &  =\left\langle \omega,b_{\sigma}\right\rangle d\beta
^{\sigma}. \label{RPS7}%
\end{align}

\item The fundamental curvature extensor field of $\left\langle
U,B\right\rangle $ vanishes, i.e., $\left\langle U,B\right\rangle $ is such
that%
\begin{equation}
\rho(a,b,c)=0. \label{RPS8}%
\end{equation}

\end{itemize}

We present only the proof of the properties given by the Eqs. (\ref{RPS6}) and
(\ref{RPS7}), the other proofs are analogous.

\begin{proof}
a) First note that the fundamental torsion extensor field, associated with the
parallelism structure $\left\langle U,B\right\rangle $, \ is defined by%
\[
\tau\left(  a,b\right)  =\partial_{a}b-\partial_{b}a-\left[  a,b\right]  ,
\]
then by using the Eq.(\ref{RPS4}), we can write
\begin{equation}%
\begin{array}
[c]{ll}%
\tau\left(  a,b\right)   & =\partial_{a}b-\partial_{b}a-\left[  a,b\right]  \\
& =\left[  a\beta^{\sigma}\left(  b\right)  \right]  b_{\sigma}-\left[
b\beta^{\sigma}\left(  a\right)  \right]  b_{\sigma}-\varepsilon^{\mu}\left(
a\right)  \varepsilon^{\nu}\left(  b\right)  c_{\mu\nu}^{\sigma}b_{\sigma}\\
& =\left[  a\beta^{\sigma}\left(  b\right)  -b\beta^{\sigma}\left(  a\right)
-\varepsilon^{\mu}\left(  a\right)  \varepsilon^{\nu}\left(  b\right)
c_{\mu\nu}^{\sigma}\right]  b_{\sigma},
\end{array}
\label{EN1}%
\end{equation}
where $\left[  a,b\right]  =\left[  \varepsilon^{\mu}\left(  a\right)  b_{\mu
},\varepsilon^{\nu}\left(  b\right)  b_{\nu}\right]  =\varepsilon^{\mu}\left(
a\right)  \varepsilon^{\nu}\left(  b\right)  \left[  b_{\mu},b_{\nu}\right]
=\varepsilon^{\mu}\left(  a\right)  \varepsilon^{\nu}\left(  b\right)
c_{\mu\nu}^{\sigma}b_{\sigma}$.

On the other hand, if $\mathcal{L}_{a}S$ denote the Lie derivative of $S$ in
the direction of $a$,  (see, e.g., \cite{choquet,rodoliv2006}), we have%
\begin{equation}%
\begin{array}
[c]{cc}%
d\beta^{\sigma}\left(  a,b\right)   & =\mathcal{L}_{a}\left(  \beta^{\sigma
}\left(  b\right)  \right)  -\mathcal{L}_{b}\left(  \beta^{\sigma}\left(
a\right)  \right)  -\varepsilon^{\sigma}\left(  \left[  a,b\right]  \right)
\\
& =a\left(  \beta^{\sigma}\left(  b\right)  \right)  -b\left(  \beta^{\sigma
}\left(  a\right)  \right)  -\varepsilon^{\mu}\left(  a\right)  \varepsilon
^{\nu}\left(  b\right)  c_{\mu\nu}^{\sigma}.
\end{array}
\label{EN2}%
\end{equation}
Thus, from the Eq. (\ref{EN1}) and (\ref{EN2}) the result follows.

b) Now, for proofing Eq. (\ref{RPS7}), note that from Eq. (\ref{RPS5}) we can
write
\[
\tau\left(  b_{\mu},b_{\nu}\right)  =\partial_{b_{\mu}}b_{\nu}-\partial
_{b_{\nu}}b_{\mu}-\left[  b_{\mu},b_{\nu}\right]  =-\left[  b_{\mu},b_{\nu
}\right]  ,
\]
and by definition we have $\Theta\left(  \omega\right)  =\frac{1}%
{2}\left\langle \omega,\tau\left(  b_{\mu},b_{\nu}\right)  \right\rangle
\beta^{\mu}\wedge\beta^{\nu}$. Then,%
\begin{equation}%
\begin{array}
[c]{ll}%
\Theta\left(  \omega\right)  & =-\frac{1}{2}\left\langle \omega,\left[
b_{\mu},b_{\nu}\right]  \right\rangle \beta^{\mu}\wedge\beta^{\nu}\\
& =-\frac{1}{2}\left\langle \omega\left(  b_{\sigma}\right)  \beta^{\sigma
},\left[  b_{\mu},b_{\nu}\right]  \right\rangle \beta^{\mu}\wedge\beta^{\nu}\\
& =\omega\left(  b_{\sigma}\right)  \left(  -\frac{1}{2}\right)  \left\langle
\beta^{\sigma},\left[  b_{\mu},b_{\nu}\right]  \right\rangle \beta^{\mu}%
\wedge\beta^{\nu}.
\end{array}
\label{EN3}%
\end{equation}
On the other hand,%
\begin{equation}%
\begin{array}
[c]{ll}%
d\beta^{\sigma} & =d\beta^{\sigma}\left(  b_{\nu}\right)  \wedge\beta^{\nu
}-\frac{1}{2}\left\langle \beta^{\sigma},\left[  b_{\mu},b_{\nu}\right]
\right\rangle \beta^{\mu}\wedge\beta^{\nu}\\
& =-\frac{1}{2}\left\langle \beta^{\sigma},\left[  b_{\mu},b_{\nu}\right]
\right\rangle \beta^{\mu}\wedge\beta^{\nu},
\end{array}
\label{EN4}%
\end{equation}
thus, from Eqs. (\ref{EN3}) and (\ref{EN4}) the result follows.
\end{proof}

\subsection{Split Theorem}

Let $\left\langle U_{0},\Gamma\right\rangle $ be a parallelism structure on
$U_{0}.$ Let us take any relative parallelism structures $\left\langle
U,B\right\rangle $ such that $U_{0}\cap U\neq\emptyset.$ There exists a smooth
$2$\emph{-covariant vector extensor field} on $U_{0}\cap U,$ namely $\gamma,$
defined by
\[
\mathcal{V}(U_{0}\cap U)\times\mathcal{V}(U_{0}\cap U)\ni(a,v)\longmapsto
\gamma(a,v)\in\mathcal{V}(U_{0}\cap U)
\]
such that
\begin{equation}
\gamma(a,v)=\beta^{\mu}(v)\nabla_{a}b_{\mu} \label{STh1}%
\end{equation}
which satisfies
\begin{equation}
\Gamma(a,v)=B(a,v)+\gamma(a,v). \label{STh2}%
\end{equation}

Such a extensor field $\gamma$ will be called the \emph{relative connection
extensor field}\footnote{The properties of the tensor field $\gamma_{\mu\nu
}^{\alpha}$ such that $\gamma(\partial_{\mu},\partial_{\nu})=\gamma_{\mu\nu
}^{\alpha}\partial_{\alpha}$ where $\{\partial_{\mu}\}$ is a basis for
$\mathcal{V}(U_{0}\cap U)$ are studied in details in \cite{rodoliv2006}.} on
$U_{0}\cap U$.

From Eq.(\ref{CDV1}),this means that for all $v\in\mathcal{V}(U_{0}\cap U):$
\begin{equation}
\nabla_{a}v=\partial_{a}v+\gamma_{a}(v),\label{STh3}%
\end{equation}
where $\nabla_{a}$ is the $a$-\emph{DCDO} associated with $\left\langle
U_{0},\Gamma\right\rangle $ and $\partial_{a}$ is the $a$-\emph{DCDO}
associated with $\left\langle U,B\right\rangle $ (note that $\gamma_{a}$ is a
smooth \emph{vector operator field }on $U_{0}\cap U$ defined by $\gamma
_{a}(v)=\gamma(a,v)$).

By using Eq.(\ref{CDF1}) and Eq.(\ref{STh3}), we get that for all $\omega
\in\mathcal{V}^{\ast}(U_{0}\cap U):$
\begin{equation}
\nabla_{a}\omega=\partial_{a}\omega-\gamma_{a}^{\bigtriangleup}(\omega),
\label{STh4}%
\end{equation}
where $\gamma_{a}^{\bigtriangleup}$ is the \emph{dual adjoint} of $\gamma_{a}$
(i.e., $\left\langle \gamma_{a}^{\bigtriangleup}(\omega),v\right\rangle
=\left\langle \omega,\gamma_{a}(v)\right\rangle $).

\subsection{Jacobian Fields}

Let $\left\{  b_{\mu},\beta^{\mu}\right\}  $ and $\left\{  b_{\mu}^{\prime
},\beta^{\mu\prime}\right\}  $ be any two pairs of dual frame fields on the
open sets $U\subseteq M$ and $U^{\prime}\subseteq M,$ respectively.

If the parallelism structure $\left\langle U,B\right\rangle $ is compatible
with the parallelism structure $\left\langle U^{\prime},B^{\prime
}\right\rangle $ (i.e., define the same connection on $U\cap U^{\prime}%
\neq\emptyset$), then we can define a smooth \textit{extensor operator
field}\emph{\ }on $U\cap U^{\prime},$ namely $J$, by%
\[
\mathcal{V}(U\cap U^{\prime})\ni v\longmapsto J(v)\in\mathcal{V}(U\cap
U^{\prime}),
\]
such that
\begin{equation}
J(v)=\beta^{\sigma}(v)b_{\sigma}^{\prime}. \label{JF1}%
\end{equation}
It will be called the \emph{Jacobian field} associated with the pairs of frame
fields $\left\{  b_{\mu},\beta^{\mu}\right\}  $ and $\left\{  b_{\mu}^{\prime
},\beta^{\mu\prime}\right\}  $ (in this order!).

Note that in accordance with the above definition the Jacobian field
associated with $\left\{  b_{\mu}^{\prime},\beta^{\mu\prime}\right\}  $ and
$\left\{  b_{\mu},\beta^{\mu}\right\}  $ is $J^{\prime}$, given by%
\[
\mathcal{V}(U\cap U^{\prime})\ni v\longmapsto J^{\prime}(v)\in\mathcal{V}%
(U\cap U^{\prime}),
\]
such that
\begin{equation}
J^{\prime}(v)=\beta^{\sigma\prime}(v)b_{\sigma}.\label{JF2}%
\end{equation}
It is the \emph{inverse extensor operator }of $J,$ i.e., $J\circ J^{\prime
}(v)=v$ and $J^{\prime}\circ J(v)=v$ for each $v\in\mathcal{V}(U\cap
U^{\prime}).$

We note that
\begin{equation}
J(b_{\mu})=b_{\mu}^{\prime}\text{ and }J^{-1}(b_{\mu}^{\prime})=b_{\mu}.
\label{JF3}%
\end{equation}

\begin{itemize}
\item Take $a\in\mathcal{V}(U\cap U^{\prime}),$ the $a$-\emph{DCDO} associated
with $\left\langle U,B\right\rangle $ and $\left\langle U^{\prime},B^{\prime
}\right\rangle ,$ namely $\partial_{a}$ and $\partial_{a}^{\prime},$ are
related by
\begin{equation}
\partial_{a}^{\prime}v=J(\partial_{a}J^{-1}(v)). \label{JF4}%
\end{equation}

\end{itemize}

\begin{proof}
A straightforward calculation, yields
\[
\partial_{a}^{\prime}v=(a\beta^{\mu\prime}(v))b_{\mu}^{\prime}=(a\beta
^{\mu\prime}(v))J(b_{\mu}).
\]
Using the identity $\beta^{\mu}(J^{-1}(v))=\beta^{\mu\prime}(v)$ (valid for
smooth extensor fields) we get
\[
\partial_{a}^{\prime}v=J((a\beta^{\mu}(J^{-1}(v))b_{\mu}),
\]
from where the expected result follows.
\end{proof}

We see that from the definition of deformed parallelism structure,
$\partial_{a}^{\prime}$ is a $J$\emph{-deformation} of $\partial_{a}.$

Then, from Eq.(\ref{DPS4}), we have%

\begin{equation}
\partial_{a}^{\prime}\omega=J^{-\bigtriangleup}(\partial_{a}J^{\bigtriangleup
}(\omega)). \label{JF5}%
\end{equation}

Finally, we note that
\begin{equation}
J^{-\bigtriangleup}(\beta^{\mu})=\beta^{\mu\prime}\text{ and }%
J^{\bigtriangleup}(\beta^{\mu\prime})=\beta^{\mu}. \label{JF6}%
\end{equation}

\section{Conclusions}

In this paper using the algebra of extensor fields developed in \cite{fmcr1}
we present a thoughtful study of the theory of a general parallelism structure
in an arbitrary real $n$-dimensional differential manifold. The highlights of
our presentation are: (i) \textit{intrinsic }versions of Cartan's first and
second structure equations for the torsion and curvature extensors which
involve the plus and minus Cartan connection operators, (ii) the concept of
deformed (symmetric) parallelism structure and the relative parallelism
structure which play, in particular, \ an important role in the understanding
of geometrical theories of the gravitational field.

\end{document}